\documentclass[aps,prb,showpacs,twocolumn,superscriptaddress]{revtex4}

\usepackage{graphicx}
\usepackage{amsmath}

\newcommand{\I}{\mathrm{i}}
\renewcommand{\k}{\mathbf{k}}
\newcommand{\q}{\mathbf{q}}
\newcommand{\mb}{\mu_{\rm B}}
\newcommand{\meff}{\mu_{\rm CW}}

\def\t2g{$t_{2g}$}
\def\eg{$e_{g}$}

\begin{document}

\title{
Magnetic fluctuations and effective magnetic moments in $\gamma$--iron
due to electronic structure peculiarities}

\author{P. A. Igoshev}
\affiliation{Institute of Metal Physics, Russian Academy of Sciences, 620990 Ekaterinburg, Russia}
\author{A. V. Efremov}
\affiliation{Institute of Metal Physics, Russian Academy of Sciences, 620990 Ekaterinburg, Russia}
\author{A. I. Poteryaev}
\affiliation{Institute of Metal Physics, Russian Academy of Sciences, 620990 Ekaterinburg, Russia}
\affiliation{Institute of Quantum Materials Science, 620107 Ekaterinburg, Russia}
\author{A. A. Katanin}
\affiliation{Institute of Metal Physics, Russian Academy of Sciences, 620990 Ekaterinburg, Russia}
\affiliation{Ural Federal University, 620002 Ekaterinburg, Russia}
\author{V. I. Anisimov}
\affiliation{Institute of Metal Physics, Russian Academy of Sciences, 620990 Ekaterinburg, Russia}
\affiliation{Ural Federal University, 620002 Ekaterinburg, Russia}

\date{\today}

\begin{abstract}
Applying the local density and dynamical mean field approximations to paramagnetic
$\gamma$-iron we revisit the problem of theoretical description of magnetic properties in a wide temperature range.
We show that contrary to $\alpha$-iron, the frequency dependence of the electronic self-energy
has a quasiparticle form for both, $t_{2g}$ and $e_g$ states.
%
In the temperature range $T=1200$--$1500$ K, where $\gamma$--iron
exist in nature, this substance can be nevertheless characterized by temperature-dependent effective local moments,
which yield relatively narrow peaks in the real part of the local magnetic susceptibility.
%
At the same time, at low temperatures $\gamma$-iron (which is realized in precipitates) is better
described in terms of itinerant picture.
%
In particular, the nesting features of the Fermi surfaces yield
maximum of the static magnetic susceptibility at the incommensurate
wave vector $\q_{\rm max}$ belonging the direction
$\q_{\rm X}-\q_{\rm W}$ ($\q_{\rm X}\equiv(2\pi/a)(1,0,0),\q_{\rm W}\equiv(2\pi/a)(1,1/2,0)$, $a$ is a lattice parameter)
in agreement with the experimental data.
%
This state is found however to compete closely with the states characterized by magnetic wave vectors
along the directions $\q_{\rm X}-\q_{\rm L}-\q_{\rm K}$, where
$\q_{\rm L}\equiv(2\pi/a)(1/2,1/2,1/2)$, $\q_{\rm K}\equiv(2\pi/a)(3/4,3/4,0)$.
From the analysis of the
uniform magnetic susceptibility we find that contrary to
$\alpha$-iron, the Curie-Weiss law is not fulfilled in a broad temperature range, although the inverse
susceptibility is nearly linear in the moderate-temperature region (1200--1500 K).
The non-linearity of the inverse
uniform magnetic susceptibility in a broader temperature range is due to the density of states peak
located close to the Fermi level.
The effective exchange integrals in the paramagnetic phase are estimated
on the base of momentum dependent susceptibility.
%
\end{abstract}


\maketitle

\section{\label{sec:intro}Introduction}

The problem of iron magnetism attracts a lot of attention till now.
Pure $\alpha$-iron has body centered cubic crystal (bcc) lattice and
it is ferromagnetic at temperatures below Curie temperature 1043~K~\cite{BH55,Binary_Alloys,Donohue_1974}.
In the temperature range between 1043 and 1183~K $\alpha$--iron is paramagnetic.
This most studied allotrope of iron becomes, however, unstable above 1183~K
because of the structural phase transition to the $\gamma$--phase~\cite{BH55,Kohlhaas_1967},
which has a face centered cubic (fcc) crystal structure~\cite{Binary_Alloys,Donohue_1974}.
The theory of the $\alpha$--$\gamma$ structural transition is still under development.
Recent investigations~\cite{Kats,Leonov,phonons,DLM} have shown an important role of
magnetic correlations for this transition.
These observations are supported by the results indicating presence of
local magnetic moments in $\alpha$-iron even above the magnetic transition temperature~\cite{Our,Tosatti}.
In view of these observations understanding of magnetic properties of $\gamma$-iron, which is on the other side
of the bcc $\leftrightarrow$ fcc transition, is of high importance.

Experimentally the temperature dependence of inverse magnetic susceptibility
in $\gamma$ phase has a very weak slope, which cannot be determined to a good accuracy
because of large spread of experimental data (see Refs.~\onlinecite{Arajs_1960,Gao_2006} and references therein).
The paramagnetic Curie temperature, extracted from a fit to experimental data is negative,
$\theta_{CW} \simeq -3451$~K, and the corresponding magnetic moment is about $\meff = 7.47 \mb$~\cite{Gao_2006}.
Therefore, the magnetic properties of $\gamma$--iron are very different from those of $\alpha$-iron,
where the paramagnetic Curie temperature is positive, $\theta_{CW} \simeq 1093$~K, and the magnetic moment
is much smaller, $\meff = 3.13 \mb$\cite{Arajs_1960}.

At low temperatures the magnetically ordered fcc phase does not exist as a single crystal due to
structural phase transition. Nevertheless magnetically ordered state can be studied
in iron precipitates in copper matrix that have the same fcc crystal structure
with slightly different lattice parameter.
The first measurements of the magnetic properties of $\gamma$--Fe precipitates were carried out
in 1960s by Abrahams {\it et al.}~\cite{Abrahams_1962}. They found it to be type--I antiferromagnet (AFM)
with small N\'{e}el temperature, $T_N = 8$~K. Later studies~\cite{Gonser_1963,Johanson_1970,Liu_1979}
showed that the N\'{e}el temperature varies between 46 and 67~K depending on
the size of iron particles in precipitates and its crystal structure which can be regarded as distorted fcc.
At the end of eighties Tsunoda and coworkers in the series of neutron scattering studies~\cite{Tsunoda,Tsunoda1,Tsunoda2,Naono}
demonstrated that the iron precipitates in cooper with truely fcc structure have a spin density wave ground state
with $\mathbf{q}\approx (2\pi/a)(1,0.127,0)$ and N\'{e}el temperature $T_N = 40$~K~\cite{Naono}.

The value of Wilson--Sommerfeld ratio, $R_{\rm W}=(\pi^2k_{\rm B}^2\chi)/(3\mb^2\gamma)$,
cannot be directly found from magnetic and calorimetric measurements
since pure $\gamma$--iron does not exist as large crystal at low temperatures.
For a rough estimation of $R_{\rm W}$ the available high-temperature value of
the uniform spin susceptibility can be used,
$\chi(T=\mbox{1000 K})\simeq50\mb^2/{\rm eV}$~\cite{Arajs_1960}.
The Sommerfeld specific heat coefficient $\gamma$ was measured for different fcc alloys
in a wide range of component concentrations\cite{Gup_1964}.
The maximal value of specific heat coefficient is in antiferromagnetic Fe:Mn alloy,
$\gamma \approx 14$ mJ/(mol$\cdot$K$^2$).
The nonmagnetic Ni:V alloy has the smallest value of specific heat coefficient,
$\gamma \approx 5$ mJ/(mol$\cdot$K$^2$).
Two above opposite limits cover the situation in the presence or absence of magnetic fluctuations
in alloys.
Therefore one finds Wilson--Sommerfeld ratio in range $8 < W_R < 25$,
which points to the presence of strong ferromagnetic fluctuations,
whether or not the magnetic contiribution to the specific heat is taken into account, and indicates that the
(antiferro)magnetism in $\gamma$--iron is likely to be frustrated by the competing magnetic fluctuations.

The ground state magnetic properties of $\gamma$-iron were considered previously
within the density functional theory calculations by many authors.
%
In the pioneering study of Mryasov {\it et al.}~\cite{Mryasov} 
the incommensurate spin spiral (SS) magnetic order was considered in the framework of
the tight-binding linearized muffin-tin orbitals with atomic sphere
approximation for the potential (TB-LMTO-ASA).
They found that for the range of lattice parameter $6.8 < a < 6.96$
the ground state energy approaches its minimum for the spiral state
with $\q=(2\pi/a)(0,0,q)$, where $q$ is close to 0.5,
while for larger lattice parameter, $a > 7.11$, the ferromagnetic state is more energetically favorable
(the atomic units are used for the lattice parameter).
Similar results were obtained within augmented spherical wave method~\cite{Uhl}.
%
Using TB-LMTO-ASA method James {\it et al.}~\cite{James} considered a stability of different magnetic structures
with increasing of the volume and found the following sequence
of magnetic phase transitions: low-spin FM $\xrightarrow{a=6.5}$ 3{\bf k} structure
$\xrightarrow{a=6.78}$ double-layered AFM $\xrightarrow{a=6.9}$ triple-layered AFM
$\xrightarrow{a=7.04}$ high-spin FM. The calculations within disordered local moments approximation
gave a metastable solution with slightly higher energy.
%
%
At the same time, spin molecular dynamics calculations, based on first-principles Kohn-Sham spectra~\cite{SMD}, applied for the $\gamma$-iron
yielded the following transitions: 2{\bf k} superimposed SS with $\q=(2\pi/a)(0,0,q) \xrightarrow{a=6.79}$
double-layered AFM $\xrightarrow{a=7.05}$ FM.
%
K\"orling and Ergon~\cite{Ergon} analyzed the importance of the full potential scheme and replacement of
the local spin density approximation by the generalized gradient one.
They found that the use of the above mentioned approximations leads to the results that
are closer to experiments than earlier studies.
Later on Kn\"opfle {\it et al.}~\cite{Knopfle} using modified augmented spherical waves
method that takes into account intra atomic magnetization non--collinearity found
that the ground state is SS with $\q \approx (2\pi/a)(0.15,0,1)$ which is close to the experimental value.
They also first noticed that 3$d$ electrons in $\gamma$-iron forms well defined local moments.
Sj\"ostedt and Nordstr\"om~\cite{Nordstrom}
demonstrated that the use of the full potential scheme with non-collinear
approach for intra atomic magnetization is more important for the proper description of the magnetic ground
state than applying different approximations for exchange correlation potential. They found the SS ground state with
the wave vector, $\q \approx (2\pi/a)(0.19,0,1)$.

One can see that quite generally the results for the type of the magnetic ground state
in $\gamma$-iron strongly depend on the value of lattice parameter and
approximations made for account of intra atomic magnetic structure and interaction potential
which may point to a close competition of different magnetic states in this material.
%
Recent analysis~\cite{Ruban,Gornostyrev} within
the ab-initio SS approach have also shown presence of long-range competing exchange interactions which strongly
depend on the lattice parameter.

The calculations of the paramagnetic state were performed within a disordered local moment approach (DLM)
by many authors~\cite{James,Abrikosov,Shallcross,Gornostyrev} who compared the stability of the paramagnetic
solution versus different SS states depending on volume.
It was found that the DLM solution lies always higher in energy with respect to the ordered state
regardless the lattice parameter value~\cite{Abrikosov,Shallcross}.
One should remember that DLM is the approach on top of density functional theory to treat
paramagnetic ground state and therefore it does not consider correlation effects.
Although the paramagnetic solution obtained with DLM can be stable at higher temperatures
its treatment requires other methods, which necessarily include correlation effects.

A possible approach for obtaining temperature evolution of magnetic properties
with account of correlation effects is a combination of local density approximation (LDA) with the
dynamical mean-field theory (DMFT). Recently, the LDA+DMFT calculations of the spectral properties and uniform magnetic susceptibility
were carried out by Pourovskii {\it et al.}~\cite{Pourovskii} for all iron allotropes.
The authors have concentrated mainly on high pressure data with small value of the volume.
They obtained that at these conditions the fcc iron is  Fermi-liquid-like material with
the exchange--enhanced Pauli susceptibility.

In the present paper we focus on the detailed LDA and LDA+DMFT calculations of magnetic susceptibilities
to investigate the origin of weak antiferromagnetism of
$\gamma$-iron, dominating types of magnetic fluctuations and possibility of the local moment formation in this substance.

\section{\label{sec:electronics}Spectral properties}

We first consider the results for $\gamma$--iron in LDA approximation.
$\gamma$-iron crystallizes in a stable face centered cubic structure in the temperature interval
from 1183~K to 1667~K and it has the lattice parameter $a =$ 6.91 a.u.~at 1183~K\cite{Binary_Alloys,Donohue_1974}.
Band structure calculations have been carried out in LDA approximation~\cite{Jones_Gunnarsson_RMP_1989}
within tight-binding linear muffin-tin orbital atomic spheres approximation
framework~\cite{Andersen_Jepsen_PRL_1984}.
The von Barth-Hedin local exchange-correlation potential has been used~\cite{Barth_Hedin_1972}.
Primitive reciprocal translation vectors have been discretized into 12 points
along each direction which leads to 72 \textbf{k}--points in irreducible part of the Brillouin zone.

\begin{figure}[tbp]
  \includegraphics[clip=true,width=0.47\textwidth]{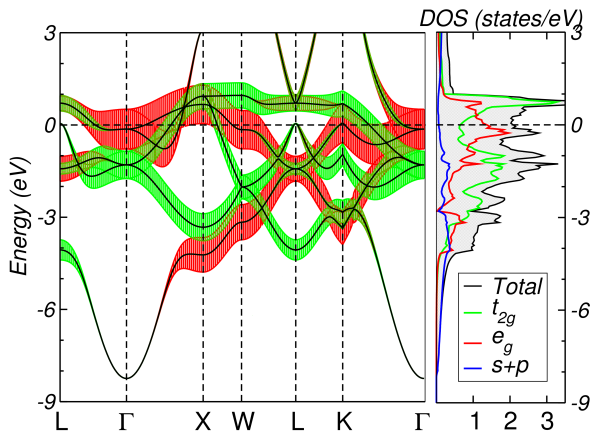}
  \caption{(Color online) Left panel: the fatbands for \t2g and \eg~ orbitals in light (green) and dark (red) gray colors, respectively.
           Fatness corresponds to appropriate partial orbital contribution.
           Right panel: Iron density of states.
           Total DOS is shown by solid (black) line.
           Partial DOSes for $t_{2g}$, $e_{g}$ and sum of $s+p$ orbitals are shown by light (green), dark (red) and dashed-dark (blue) gray lines, respectively.}
  \label{fig:dos}
\end{figure}

The band structure together with the density of states are presented in the Fig.~\ref{fig:dos}.
On the left part of the figure the fatbands for the \t2g and \eg~ orbitals are shown by green and
red colors, respectively (light and dark gray in the black-and-white version).
The fatness coincides with the contribution of the corresponding partial DOSes
shown on the right part of the Fig.~\ref{fig:dos}.
The bands of $t_{2g}$ and $e_g$ symmetries hybridize in the vicinity of
the L point and in K$-\Gamma$ direction.
In other symmetry directions the \t2g and \eg~ manifolds hybridize weakly with $s$ and $p$ bands which span energy range from -8~eV to far above Fermi level (corresponding to zero energy).
The \t2g states have a very flat region along X$-$W$-$L$-$K directions that is reflected in the DOS peak at 0.7~eV. At the Fermi level the partial \t2g DOS has a deep.
Other large peaks of the \t2g DOS are located at $-1.3$ and $-2.6$~eV.

Although the \eg~ partial DOS has a bandwidth almost equal to the \t2g counterpart, its shape is very different.
The corresponding dispersion has a flat part at small negative energy near $\Gamma$ point
(extended van Hove singularity, cf. Ref.~\onlinecite{Vonsovskii_FMM_1993}), which
results in the large peak of DOS just below the Fermi level at about $-0.2$~eV,
such that the states at the Fermi energy lie at the slope of peak.
The smaller peak of the corresponding partial DOS is located at $-3.4$~eV.
This is in contrast to $\alpha$-iron~\cite{Maglic_PRL_1973, Our}, where peak of $e_g$ density of states is located
very close to the Fermi level.
As it will be shown below, this shift is of crucial importance for the magnetic properties difference
between $\alpha$-- and $\gamma$--iron.

\begin{figure}[tbp]
	\includegraphics[clip=true,width=0.23\textwidth,angle=270,keepaspectratio]{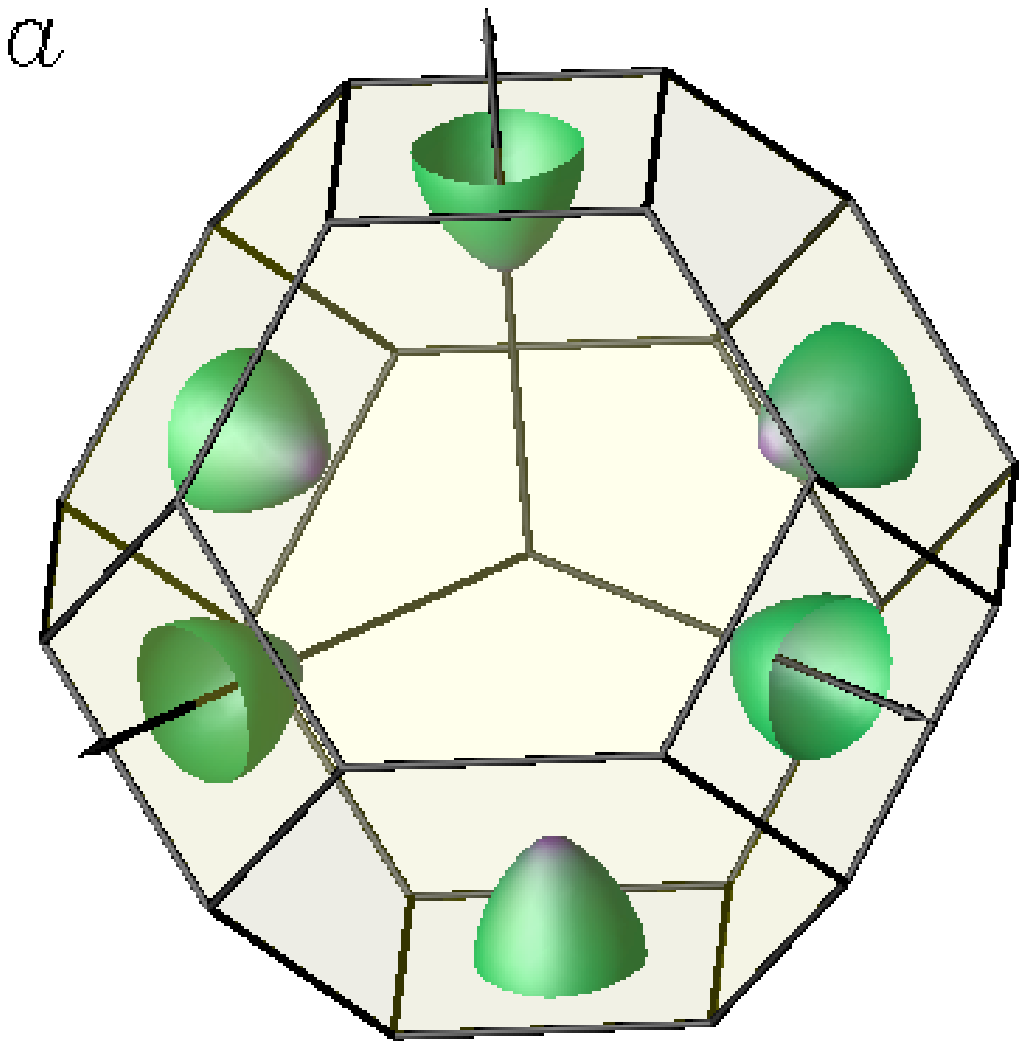}
        \includegraphics[clip=true,width=0.23\textwidth,angle=270,keepaspectratio]{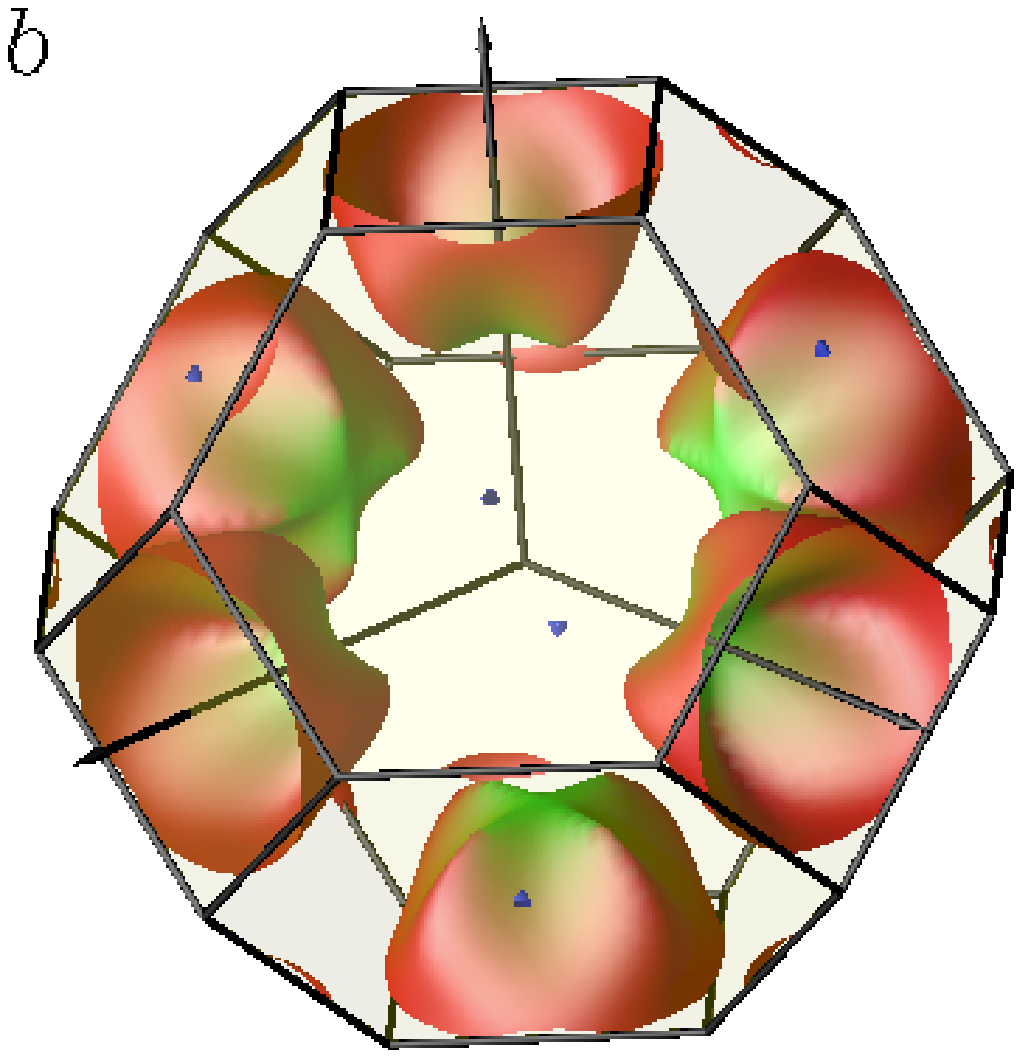}
        \includegraphics[clip=true,width=0.23\textwidth,angle=270,keepaspectratio]{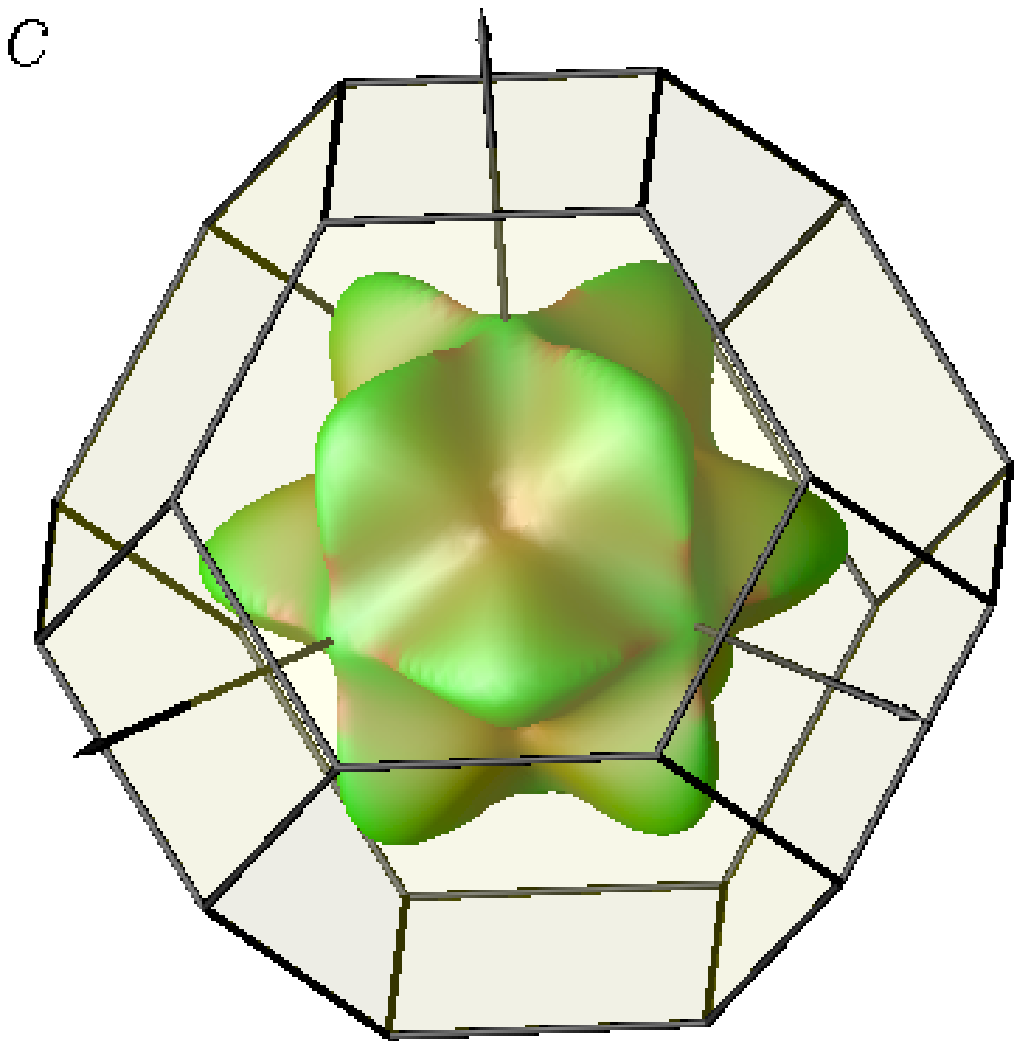}
        \includegraphics[clip=true,width=0.23\textwidth,angle=270,keepaspectratio]{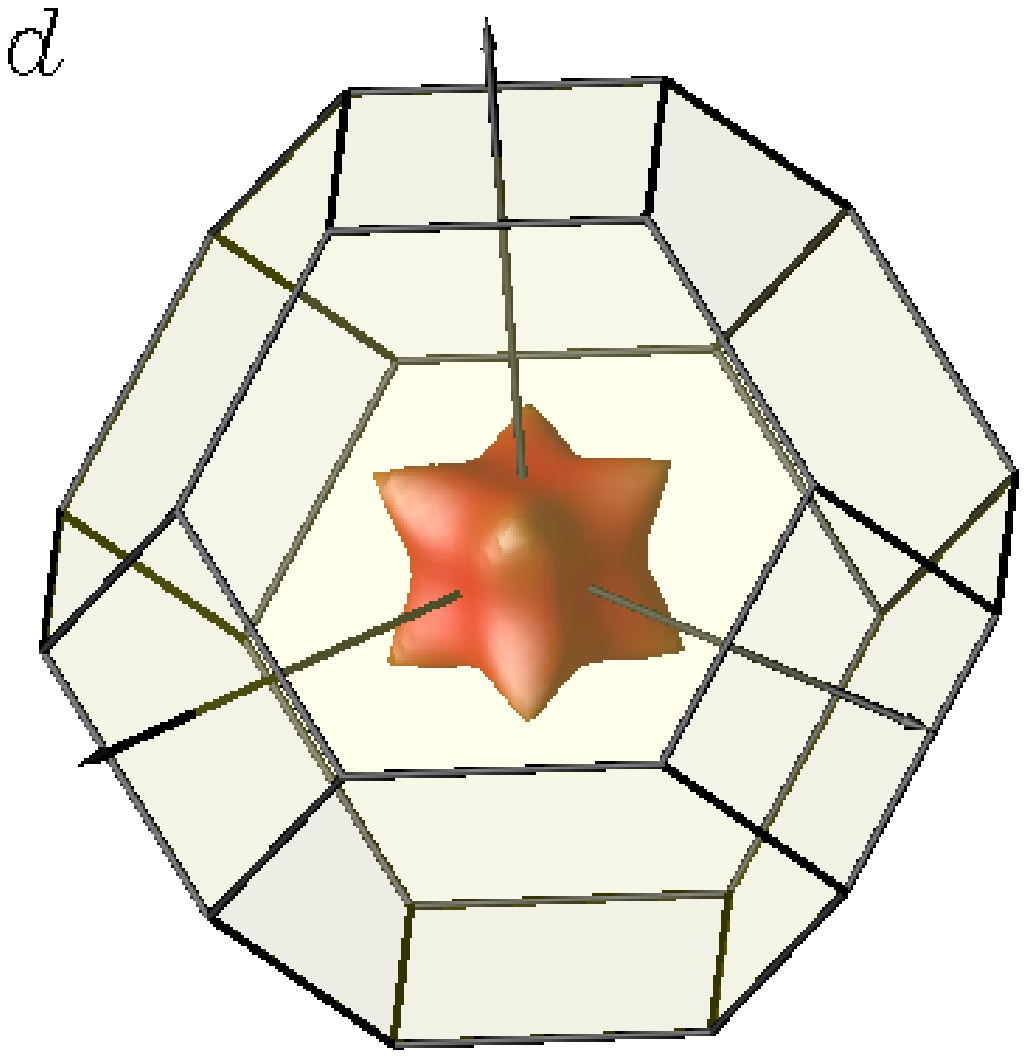}
	\caption{(Color online) $\gamma$-iron Fermi surface sheets.
		The colorcoding reflects contribution of the orbital states.
		The (red,green,blue) scheme is used for the color definition of the point
		where red is for \eg, green is for \t2g and blue is for $s+p$ orbitals, respectively.}
	\label{fig:FS}
\end{figure}

The Fermi surface obtained within LDA is shown in the Fig.~\ref{fig:FS}.
The four sheets that satisfy the equation for the Fermi surface, $\varepsilon _{\mathbf{k}_F} = 0$,
are colored such that amount of the appropriate color corresponds to the weight of partial contribution
(we use the same colors as in Fig. 1: red for \eg~ states, green for \t2g states and blue for $s+p$ orbitals, respectively).
The sheet \emph{a} of the Fermi surface (Fig.~\ref{fig:FS}a) is of mostly $s$, $p$, and \eg~ orbital characters.
The sheets \emph{b} and \emph{c} (Figs.~\ref{fig:FS}b,c) are mixture of \t2g and \eg~ characters.
The last sheet \emph{d} (Fig.~\ref{fig:FS}d) consists mostly of \eg~ states.
One should note that \emph{b} and \emph{c} sheets touch each other at the wavevector $(2\pi/a)(0.57,0,0)$ and
thus lead to the three bands crossing the Fermi level along $\Gamma-$X direction (see Fig.~\ref{fig:dos}).
Near the touch point these sheets have a cross-like features  with the small opposite incurvature
perpendicular to [0,0,1] direction produced by mostly \t2g states.
This results in the approximate interband nesting of these crossed parts
with close to zero wavevector and the intraband nesting with the wavevector $\q_{\rm A}=(2\pi/a)(0.86,0,0)$.
The \emph{d }sheet reminds the cube stretched along diagonals and it has also the cross-like feature.
Its existence allows one to consider two additional candidates for nesting vectors:
within this sheet with $\q_{\rm B}=(2\pi/a)(0.48,0,0)$ and the vector connecting the sheets \emph{b,c} and \emph{d},
$\q_{\rm C}=(2\pi/a)(0.81,0,0)$.



In order to take into account correlation effects in 3$d$ shell of $\gamma$--iron we apply the LDA+DMFT method
(for a detailed description of the computation scheme see Refs.~\onlinecite{Anisimov_2005,Lechermann}).
The Coulomb interaction parameter value, $U$ = 2.3~eV, and the Hund's parameter, $I$ = 0.9~eV,
used in our work are the same as in earlier LDA+DMFT calculations by
Lichtenstein {\em et al.}~\cite{Lichtenstein_2001} for $\alpha$--iron.
The effective impurity model for DMFT was solved by quantum-Monte-Carlo (QMC) method
with the Hirsch-Fye algorithm~\cite{Hirsch_Fye_1986}.
Calculations were performed for the value of temperature $T \approx$ 1290~K which is just
above the $\alpha$--$\gamma$ structural transition temperature.
Inverse temperature interval $0 < \tau < \beta\equiv 1/k_BT$ was divided in 100 slices.
Four million QMC measurements were used in self-consistency loop within LDA+DMFT scheme and
up to twelve million to refine data for spectral functions calculation with maximum entropy method~\cite{Sandvik}.
We also consider room temperature
$T=290$ K within the CT--QMC algorithm, adopting the lattice parameter to the value $a =$ 6.75~a.u.,
which is found by linear extrapolation of the experimental data to the considered temperature.

\begin{figure}[tbp]
  \includegraphics[clip=true,width=0.33\textwidth,angle=270]{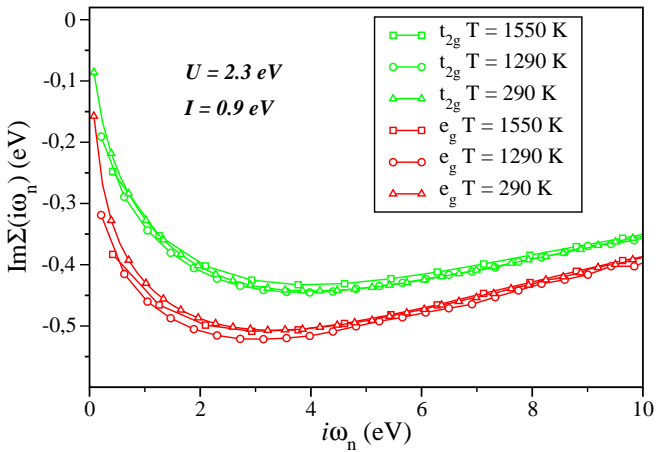}
  \caption{(Color online) The imaginary parts of self--energies for \t2g (green in color) and \eg ~states (red in color), lattice parameter $a=3.656$ \AA, plotted on the
Matsubara energy grid for different temperatures ($T$=1290~K --- circles, $T=1550$~K --- squares, and $T=290$~K --- triangles).}
  \label{fig:sigma}
\end{figure}

\begin{figure}[tbp]
  \includegraphics[clip=true,width=0.47\textwidth]{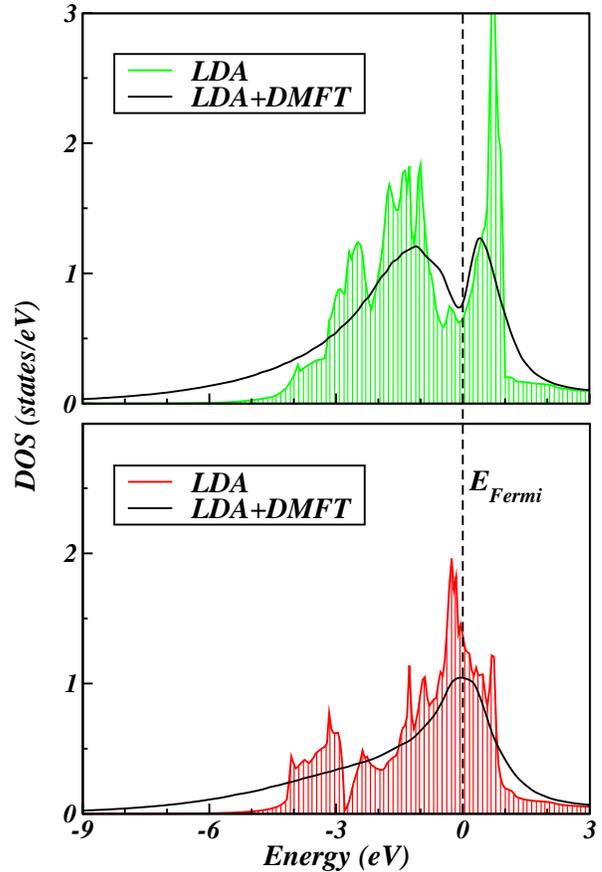}
  \caption{(Color online) The $t_{2g}$ (top panel) and $e_{g}$ (bottom panel)
            partial density of states of $\gamma$--iron,
            obtained within LDA (filled) and LDA+DMFT method (solid lines).}
  \label{fig:dmft_dos}
\end{figure}

The imaginary parts of self-energies for $a =$ 6.91 a.u. are presented in the Fig.~\ref{fig:sigma}
(the results for smaller lattice parameter, $a =$ 6.75 a.u., are qualitatively similar).
At low energies the behavior of the $\Im \Sigma(i\omega_n)$ is qualitatively similar for the \t2g and \eg~orbitals.
One can clearly see that increase of temperature does not change the frequency dependence qualitatively.
The effective mass stays close to the bare value, $m^*/m\lesssim1.2$, and increases slightly
in temperature interval 1220~K $< T <$ 1550~K,
where $\gamma$--iron exists in nature. The damping of electronic states also increases with increasing temperature,
especially for $e_{g}$ states.
However, the obtained imaginary part of \eg~self-energy in $\gamma$--Fe has a quasiparticle-like frequency dependence
at all considered temperatures, in stark contrast with the non-quasiparticle frequency dependence in $\alpha$--phase~\cite{Our}.
The reason of this difference between $\gamma$-- and $\alpha$--iron seems to lie in the shift of the DOS peak
from the Fermi level in $\gamma$--iron.
We would like to note that the shift of the peak of the density of
states also yields more quasiparticle self-energies in iron-based superconductors~\cite{Pniktides}.

The LDA+DMFT densities of states in $\gamma$--iron (see Fig. \ref{fig:dmft_dos}) are slightly narrower than the LDA counterparts implying weak correlation effects.
This is in agreement with the small mass renormalization.
One can observe that peak of $e_{g}$ density of states obtained in LDA approach is broadened in LDA+DMFT calculation.
This is in contrast to $\alpha$-iron, where the density of states, corresponding to $e_g$ orbitals, is strongly renormalized by
the interaction.
The shape of \t2g density of states in LDA+DMFT approach resembles the LDA result with smearing of the peaky structures
in both, $\alpha$-- and $\gamma$--iron.

\begin{figure}[tbp]
	\includegraphics[clip=true,width=0.35\textwidth,angle=270]{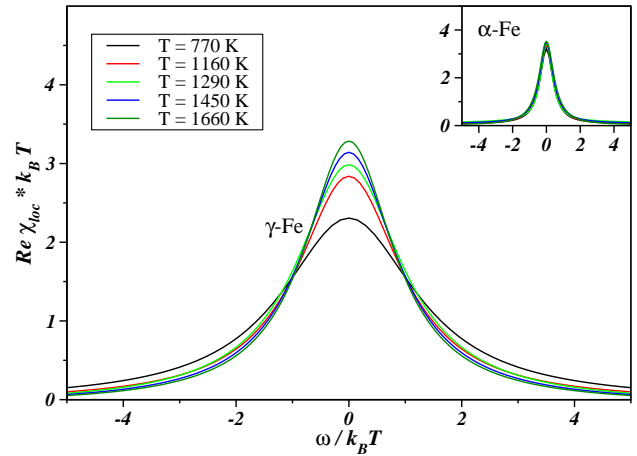}
	\caption{(Color online) Local magnetic susceptibility of $\gamma$--iron for different temperatures.
	         The inset shows the results for $\alpha$--iron.}
	\label{fig:chi_local}
\end{figure}

To investigate the possibility of the local moment formation in $\gamma$--iron,
the analytic continuation of the dynamic local magnetic susceptibility
\begin{equation}
\chi
_{\text{loc}}(i\omega_{n})=\mb^2\int_0^{\beta} d\tau\langle S_i^{z}(0)S_i^{z}(\tau)\rangle
e^{i\omega_{n}\tau}
\end{equation}
(where ${\bf S}_i=\sum _{m \sigma \sigma ^{\prime }}%
\hat{c}_{im\sigma }^{\dagger }\boldsymbol{\sigma}_{\sigma \sigma ^{\prime }}%
\hat{c}_{im\sigma ^{\prime }}$, $\hat{c}_{im\sigma }^{\dagger }$,$\hat{c}_{im\sigma }$ are
the electron creation and destruction operators
at a site $i$, orbital $m$, and spin projection $\sigma$,
$\boldsymbol{\sigma} _{\sigma \sigma ^{\prime }}$ are the Pauli matrices)
to real frequency axis have been calculated.
In Fig.~\ref{fig:chi_local} we present real parts of the obtained functions for different temperatures,
rescaling both the susceptibility and frequency by temperature.
For comparison, we also present on the
inset the corresponding result for $\alpha$-iron (see also Ref.~\onlinecite{Our}).

The results for the low-energy behavior of $\chi_{\rm loc}(\omega)$, in both $\alpha$- and $\gamma$--iron,
can be well fitted by the simple form
\begin{equation}
\chi_{\text{loc}}(\omega)=\frac{\mu_{\rm eff}^{2}}{3T}\frac{i\delta}%
{\omega+i\delta}\label{hi_loc}%
\end{equation}
yielding Lorentzian frequency dependence of $\Re\chi_{\text{loc}}$ with
$\delta$ corresponding to a halfwidth of its peak
at a half--height (or, equivalently, to the position of the maximum of $\Im\chi_{\text{loc}}(\omega)$).
In the Eq. (\ref{hi_loc}) we have picked out factor \thinspace$1/T$ to emphasize the expected Curie law of the static
susceptibility in the local-moment regime, $\chi_{\rm loc}\equiv \chi_{\text{loc}}(0)=\mu_{\rm eff}^{2}/(3T)$, while in general the effective moment
$\mu_{\rm eff}$ is temperature-dependent.
The Eq. (\ref{hi_loc}) implies that
the width $\delta$ of the peak of $\Re\chi_{\text{loc}}$
 describes the damping of local excitations (or
their inverse lifetime).
For $\alpha$ iron we find $\delta$ is linear with temperature, $\delta\simeq T/2$
for $T<1200$ K,
while in the temperature range, where $\gamma$--iron exist in nature, we obtain $\delta\simeq(1\div1.5)T,$ which implies
smaller life time of the local moments;
for lower
temperatures we obtain even bigger values $\delta>2T$.

For the system with the local moments the dynamical mean-field theory, which neglects intersite magnetic exchange
and therefore has no other low-energy scales
apart from temperature, is expected to yield the low-frequency
part of the local magnetic susceptibility in the form
$\chi_{\rm loc}(\omega)=(1/T)f(\omega/T)$,
with some function $f(x)$ which tends to zero at $x\rightarrow \infty$.
Such a dependence for the Eq. (\ref{hi_loc}) implies $\delta \propto T$ and $\mu_{\rm eff}$ is temperature-independent,
which
naturally provides the static nature of a single spin,
$\chi_{\rm loc}\propto \delta(\omega)$ at $T\rightarrow 0$. This dependence
agrees with obtained results for $\alpha$-iron,
while for $\gamma$--iron some deviations are observed.

%
%

\begin{figure}[tbp]
	\includegraphics[clip=true,width=0.45\textwidth]{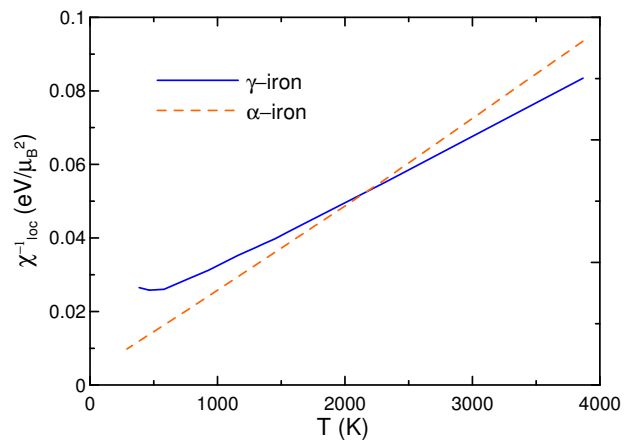}
	\caption{(Color online) Temperature dependence of the inverse static local magnetic susceptibility of $\alpha$--
             and $\gamma$--iron}
	\label{fig:chi_local_vs_T}
\end{figure}

\begin{figure}[tbp]
	\includegraphics[clip=true,width=0.45\textwidth]{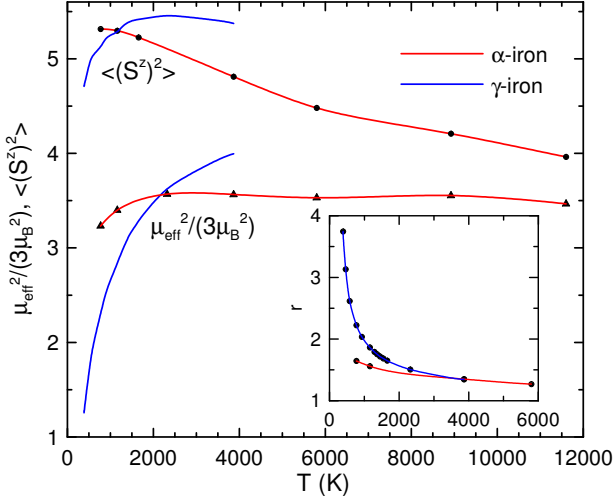}
	\caption{(Color online) The temperature dependence of the effective magnetic moment and
	         instantaneous average $\langle (S^z)^2 \rangle$ in $\alpha$-- and $\gamma$--iron, extracted from the frequency dependence of
	         local susceptibility, see Eq. (\ref{hi_loc}). Inset shows the temperature dependence of the ratio
                 $r=3 \mb^2 \langle (S^z)^2 \rangle/\mu_{\rm eff}^2$}
	\label{Fig:mu_eff}
\end{figure}

The inverse static local magnetic susceptibility, $\chi_{\rm loc}$, is shown
on Fig.~\ref{fig:chi_local_vs_T}. One can see that for both, $\alpha$- and $\gamma$-iron the inverse static local susceptibility
is almost linear with temperature in a broad temperature range
with some non-linearity at the low temperatures for $\gamma$-iron.
In the linear regime
the inverse local susceptibility fulfills the dependence $\chi_{\rm loc}^{-1}\approx 3(T+\Theta)/\mu_{\rm loc}^2$,
which has a constant part proportional to the temperature $\Theta$,
appearing due to local fluctuations; fitting the obtained temperature dependences we obtain for $\gamma$-iron $\mu_{\rm loc}\approx 3.8 \mb$
(corresponding to the spin $S\approx 3/2$) and $\Theta \approx 800$ K, while for $\alpha$-iron $\mu_{\rm loc}\approx 3.13 \mb$
(corresponding to the spin $S\approx 1.15$) and $\Theta \approx 100$ K.
The temperature dependence of $\chi_{\rm loc}$ provides peculiarities of the temperature dependence of $\mu _{\rm eff}$,
which is shown on the Fig.~\ref{Fig:mu_eff}. This dependence approximately fulfills
\[
\mu _{\rm eff}\approx \mu _{\text{loc}}
\sqrt{T/(T+\Theta )}.
\]
At $T\gg \Theta$ (which is fulfilled for realistic temperatures for $\alpha$-iron only) the size of the effective moment slightly varies with temperature,
while in $\gamma$-iron
we find a variation of $\mu_{\rm eff}$ with temperature, which is mainly due to above mentioned constant contribution
in the inverse susceptibility.
In the temperature region 1200--1400~K we obtain for $\gamma$-iron $\mu_{\rm eff}\approx 3 \mb$.

The obtained temperature dependence of instantaneous average $\langle(S^{z})^{2}\rangle$
is qualitatively similar to that of $\mu_{\rm eff}^2$,
although the former quantity does not remain approximately constant even for $\alpha$-iron (see Fig.~\ref{Fig:mu_eff}).
Considering the ratio $r=3\langle(S^{z})^{2}\rangle/\mu_{\rm eff}^2$, shown in the inset of Fig.~\ref{Fig:mu_eff},
we see however that for $\alpha$-iron $r$ is of the order of 1 in a broad temperature range.
As it is shown in Appendix, this requires
$\delta \ll \pi T$, which is well fulfilled for $\alpha$-iron. Accepting the latter criterion as a condition of
the existence of sufficiently long-living local moments, we find that for
$\gamma$-iron it is fulfilled
only at the intermediate and high-temperatures $T>1000$ K (where $r$ also approaches values of the order of 1),
indicating possible local nature of electronic states in that limit.
This conclusion also agrees with the linear dependence of $\chi_{\rm loc}^{-1}$ in the above discussed temperature range.
At low temperatures the criterion $\delta \ll \pi T$ is violated for $\gamma$-iron, and $r$ increases to
the values much larger than one,
showing that the local moments in $\gamma$-iron at low temperatures are not well defined,
which is also consistent with the quasiparticle form of the self-energy.

%


\section{\label{sec:magnetism}Magnetic properties}

To gain insight into favorability of different types of magnetic order in $\gamma$--iron,
we analyze the momentum $\bf q$-dependence of generalized static magnetic susceptibility
$\chi_{\q}$ within LDA and LDA+DMFT approximations.
The static magnetic susceptibility without correlation effects can be obtained as
\begin{eqnarray}
  \label{eq:chi_LDA}
  \chi_{\q}^{0} &=&
\mb^2\int_0^{\beta} d\tau\langle S_i^{z}(0)S_j^{z}(\tau)\rangle e^{i {\bf q} ({\bf R}_i-{\bf R}_j)} \notag \\
&=&
- \frac{2\mb^2}{\beta}\sum_{{\k},\omega_n}{\rm Tr} \left[ \mathcal{G}^{\rm LDA}_{\k}(\I\omega_n)
                         \mathcal{G}^{\rm LDA}_{\k+\q}(\I\omega_n) \right],
\end{eqnarray}
where the Green function $\mathcal{G}^{\rm LDA}_{\k}(\I\omega_n)=(\I\omega_n-\mathcal{H}_{\k}+\mu)^{-1}$, $\mu$ is the chemical potential and $\mathcal{H}_{\k}$ is the LDA--constructed Hamiltonian.
Note that the temperature in Eq.~(\ref{eq:chi_LDA}) is introduced via the Fermi distribution function only.
To analyze the contribution of  different orbitals to the susceptibility, we represent Green function
\begin{equation}
  \mathcal{G}^{\rm LDA}_{\k}(\I\omega_n) = \sum_{\alpha m_1 m_2} |m_1\rangle
       \frac{\bar\psi^{\alpha m_1}_\k \psi^{\alpha m_2}_\k }{\I\omega_n-\varepsilon_{\alpha\k}}
                          \langle m_2|,
\end{equation}
where $\{|m\rangle\}$ is an orbital (LMTO) basis and $\psi^{\alpha m}_\k$ ($\varepsilon_{\alpha\k}$) are LDA eigenvectors (eigenvalues) written in orbital representation ($\alpha$ is a band index).
In this notation the equation~(\ref{eq:chi_LDA}) can be rewritten as
\begin{eqnarray}
  \chi_{\q}^{0} &=& -\frac{2\mb^2}{\beta} \sum_{\k n} \sum_{\alpha_1,\alpha_2\atop m_1,m_2}
        \frac{\bar\psi^{\alpha_1m_1}_\k\psi^{\alpha_1m_2}_\k\bar\psi^{\alpha_2m_2}_{\k+\q}\psi^{\alpha_2m_1}_{\k+\q}}%
        {(\I\omega_n-\varepsilon_{\alpha_1\k})(\I\omega_n-\varepsilon_{\alpha_2\k+\q})}\notag \\
&=& \chi_{\q}^{0,d}+\chi_{\q}^{0,{\rm rest}}. \label{eq:detail_contrib}
\end{eqnarray}
where $\chi_{\q}^{0,d}$, corresponds to restricting the $m_{1,2}$ sum
over $d$-orbitals only, while $\chi_{\q}^{0,{\rm rest}}$, contains the rest.
For the following analysis we also split the susceptibility according to the contribution of different orbotals:
\begin{equation}
  \chi^{0,d}_{\q} = \chi^{0,e_{\rm g}-e_{\rm g}}_{\q}    +
                    \chi^{0,t_{\rm 2g}-t_{\rm 2g}}_{\q}  +
                    \chi^{0,e_{\rm g}-t_{\rm 2g}}_{\q}.
  \label{eq:decomp}
\end{equation}

\begin{figure}[tbp]
  \includegraphics[clip=true,width=0.47\textwidth]{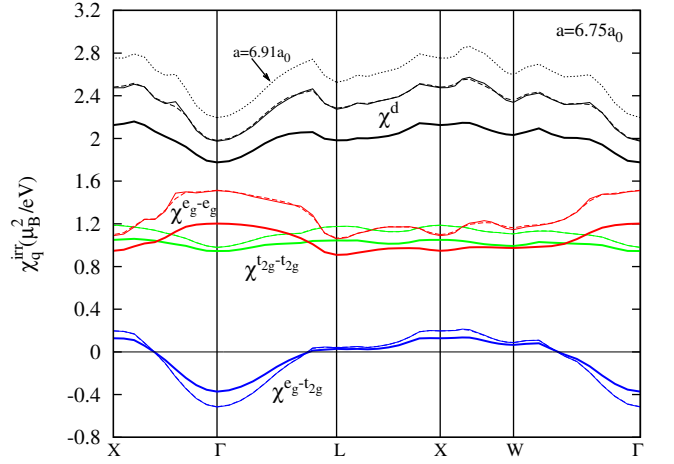}
  \caption{(Color online) Contributions of different orbitals to magnetic susceptibility calculated
           along high symmetry directions at $a=$ 6.75~a.u.
           LDA results (see Eq.~\ref{eq:decomp}) are shown by dashed and thin solid lines
           for $T=0$ K and $T=290$ K, respectively.
           LDA+DMFT data (Eq.~\ref{eq:chi_irr}) are presented by thick solid lines for $T=290$ K.
           Black line corresponds to $\chi^{0,d}_{\q}$.
           Red, green and blue lines show $\chi^{0,e_{\rm g}-e_{\rm g}}_{\q}$,
           $\chi^{0,t_{\rm 2g}-t_{\rm 2g}}_{\q}$ and $\chi^{0,e_{\rm g}-t_{\rm 2g}}_{\q}$, respectively.
           $\chi^{0,d}_{\q}$ contribution for larger lattice parameter, $a=$ 6.91~a.u., and $T=0$ is shown by thin dotted line.
           }
  \label{fig:lda_chi}
\end{figure}
The results of calculation of different contributions to the non-uniform magnetic susceptibility are presented in Fig.~\ref{fig:lda_chi}
for $a=6.75$~a.u. and sufficiently low temperatures.
The maximum of the resulting susceptibility $\chi_{\bf q}^{0,d}$ is obtained
in $\q_{\rm X}-\q_{\rm W}$ direction ($\q_{\rm X}\equiv(2\pi/a)(1,0,0)$, $\q_{\rm W}\equiv(2\pi/a)(1,1/2,0)$)
at the wavevector ${\bf q}_{\rm max}\approx (2\pi/a)(1,0.2,0)$,
which is close to results of low-temperature measurements of Tsunoda~\cite{Tsunoda}
and previous band-structure calculations~\cite{Hirai}.
Note that the change of lattice parameter to $a=6.91$~a.u. (thin dotted line) does not change
the results qualitatively, only rescaling them.
%

Considering the decomposition of the susceptibility according to the Eq. (\ref{eq:decomp}), we find that
the intra--orbital contributions to the susceptibility at zero temperature, $\chi^{0,e_{\rm g}-e_{\rm g}}_{\q}$ and
$\chi^{0,t_{\rm 2g}-t_{\rm 2g}}_{\q}$, are of the same magnitude and varying in ``counter--phase'' and thus compensating
partly the $\q$ dependence of each other. The $e_g-e_g$ contribution has a broad peak centered at the point
$\q_\Gamma=(0,0,0)$,
favoring ferromagnetic ordering, containing also features at the nesting wavevectors  $\q_{\rm B}$ and $\q_{\rm C}$,
discussed in Sec.~\ref{sec:electronics}, and two smaller peaks in the $\q_{\rm X}-\q_{\rm W}$ and $\q_{\rm X}-\q_{\rm L}$
directions ($\q_{\rm L}\equiv(2\pi/a)(1/2,1/2,1/2)$), which seem to occur due to partial nesting between sheets \emph{b}
of the Fermi surface.
Note that the momentum dependence of $e_{\rm g}$--$e_{\rm g}$ contribution
is much stronger affected by the temperature than
that of $t_{\rm 2g}$--$t_{\rm 2g}$ and $t_{\rm 2g}$--$e_{\rm g}$,
which is due to peculiarities of the $e_g$ band dispersion in the vicinity of the Fermi level, in particular
small size and cubic--corner--like form of \emph{d} sheet of the Fermi surface, and also
flatness of the corresponding electronic spectrum along the direction $\Gamma-$L.
The momentum dependence of $t_{2g}-t_{2g}$ contribution is weaker and has maxima at wavevectors $\q_{\rm X}$
and $\q_{\rm L}$, which are
related to the intraband nesting of the \emph{c} Fermi surface sheet.
The large part of the momentum dependence of susceptibility comes from $e_g-t_{2g}$ contribution,
which, at zero temperature, has a weak maximum approximately in the center of $\q_{\rm X}$--$\q_{\rm W}$ direction,
 occuring because of nesting features of \emph{c} and \emph{d} sheets of the Fermi surface, 
and negative and large by magnitude in the vicinity of ${\bf q}=0$ point
due to small momentum transfer between electron-like (mainly \t2g-derived)
Fermi-surface sheet \emph{c} and hole-like (mainly \eg-derived) sheet \emph{b}.





The effects of electron-electron interaction can be treated within LDA+DMFT approach.
Since, in general, interaction produces vertex corrections to a single bubble considered above,
we neglect for sake of simplicity the frequency dependence
of these vertex corrections, introducing the frequency-independent vertex $\Gamma^{\rm irr}$, such that
\begin{equation}
  \label{eq:chi_DMFT}
  (\chi_{\q}^0)^{-1} \rightarrow (\chi_{\q})^{-1} = (\chi_{\q}^{\rm irr})^{-1} - \Gamma^{\rm irr},
\end{equation}
where
\begin{equation}
  \label{eq:chi_irr}
  \chi_{\q}^{\rm irr} = - \frac{2\mb^2}{\beta}\sum_{n,{\k}}{\rm Tr}
      \left[\mathcal{G}^{\rm DMFT}_{\k}(\I\omega_n)\mathcal{G}^{\rm DMFT}_{\k+\q}(\I\omega_n)\right],
\end{equation}
and
\begin{equation}
  \left({\mathcal{G}^{\rm DMFT}_{\k}}(\I\omega_n)\right)^{-1} =
  \left({\mathcal{G}^{\rm LDA}_{\k}}(\I\omega_n)\right)^{-1}  -
   \mathcal{P}_d\Sigma(\I\omega_n)\mathcal{P}_d+\delta\mu.
\end{equation}
$\Sigma(\I\omega_n)$ is DMFT self-energy with subtracted double counting term,
$\mathcal{P}_d$ is a projector onto $d$-orbitals and $\delta\mu$ is a change of the chemical potential
in DMFT with respect to LDA value.

$\q$--dependence of orbitally-resolved contributions in high symmetry directions of Brillouin
zone to the irreducible susceptibility in LDA+DMFT approach are presented in the Fig.~\ref{fig:lda_chi}. One
can see that the DMFT self-energy corrections lead to suppression of irreducible susceptibility,
not changing qualitatively its momentum dependence. The latter agrees with the quasiparticle form
of the self-energy at low temperatures.

\begin{figure}[tbp]
  \includegraphics[width=0.47\textwidth]{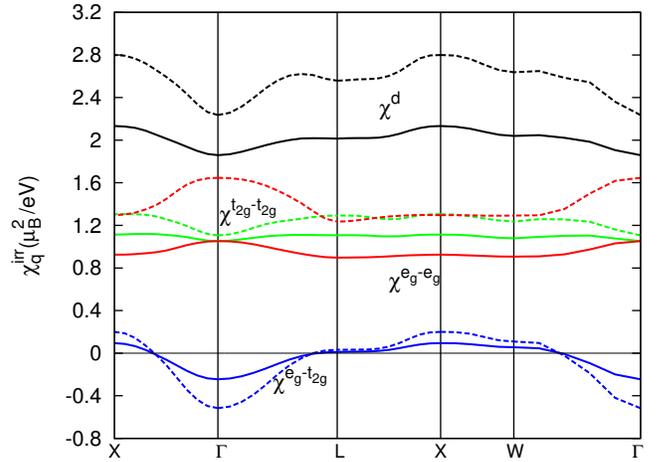}
  \caption{(Color online) Contributions of different orbitals to irreducible susceptibility calculated according to the Eq.~(\ref{eq:chi_LDA} (dashed lines) and in LDA+DMFT approach (Eq.~(\ref{eq:chi_irr}), solid lines) in high symmetry directions at $T=$1290~K, $a=6.91$~a.u.
	Colorcoding and units repeats the previous picture.}
  \label{fig:chi:LDA_vs_DMFT}
\end{figure}

Increase of temperature up to $T=1290$ K and corresponding increase of lattice parameter to $a=6.91$~a.u.
(corresponding to the thermal expansion, see Ref.~\onlinecite{lattice_a}) smears the local maximum of
$\chi^{0,e_{\rm g}-e_{\rm g}}_{\bf q}$ in the $\q_{\rm X}$--$\q_{\rm W}$ direction and
makes the corresponding momentum dependence in this direction almost flat (see Fig. \ref{fig:chi:LDA_vs_DMFT}).
The maximum of $e_g-t_{2g}$ contribution is shifted, together with the maximum of the $d$-orbital susceptibility to the
wave vector $\q_{\rm X}$, stabilizing even further the antiferromagnetic fluctuations.
The wave vector $\q_{\rm X}$ corresponds to antiferromagnetic structure with alternating orientation of magnetic
moments in adjacent layers of fcc crystal structure. We note that these effects are
mainly due to change of temperature; the lattice parameter yields only small quantitative changes of the momentum
dependence of the susceptibility. This result is not changed if one considers the increasing temperature
without the account of lattice expansion (not shown in the figure).
The flat region implies close competition of the antiferromagnetic fluctuations
with the wavevectors along the directions $\q_{\rm X}-\q_{\rm L}-\q_{\rm K}$ ($\q_{\rm K}\equiv(2\pi/a)(3/4,3/4,0)$).
According to the general ideas of spin-fluctuation theory~\cite{Moriya}, the
weak momentum dependence of the irreducible susceptibility can be also attributed to the partial presence of local
moments.


\begin{figure}[tbp]
	\includegraphics[width=0.47\textwidth]{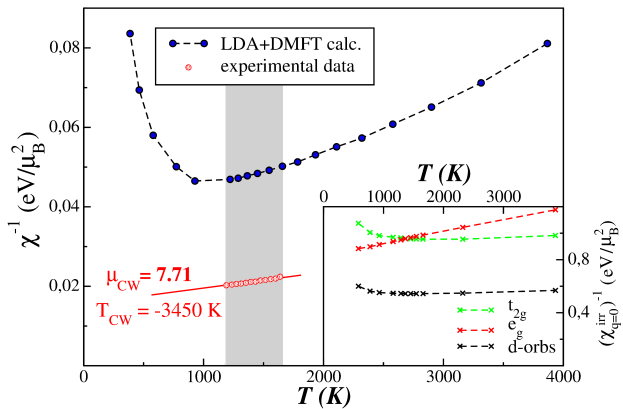}
	\caption{(Color online) Temperature dependence of the inverse uniform magnetic susceptibility
	          calculated within LDA+DMFT (blue circles) and experimental data~\cite{Arajs_1960}(red circles),
	          red line corresponds to the least square fit to Curie-Weiss law.
	          Shadow covers the temperature range of $\gamma$-phase existence.
	          Inset shows the inverse total (black) and orbital (red--$e_g$, green--$t_{2g}$) contributions to $\chi^{\rm irr}_{\q=0}$.}
	\label{fig:chi_uniform}
\end{figure}

To get further insight into the interplay of different magnetic fluctuations in $\gamma$-iron,
we consider the uniform magnetic susceptibility; the latter can give a key for understanding
the role of magnetic fluctuations.
The uniform magnetic susceptibility $\chi(T)$ in the paramagnetic state of
$\gamma$-iron was extracted from the LDA+DMFT
simulations as a ratio of the induced magnetic moment by a small external magnetic field and
the field magnitude~\cite{Pniktides,VO2}. The temperature dependence of $\chi^{-1}(T)$ is presented on
Fig.~\ref{fig:chi_uniform}. We note the absence of fulfillment of the Curie-Weiss law
\begin{equation}
\chi(T)=\frac{\meff^2}{3(T-\theta_{\rm CW})},
\end{equation}
up to highest considered temperatures, in contrast to the local susceptibility, analyzed in Sect. II.
The uniform inverse susceptibility $\chi^{-1}(T)$ has a well pronounced minimum at $T^* \simeq 1000$ K,
related to the presence of the peak of the density of states near the Fermi level, as discussed below.

The effective magnetic moment, extracted from
the slope of the inverse susceptibility in the temperature region 1200--1550 K,
$\meff$ = 5.75$\mb$, is close to the experimentally observed value,
$\meff=7.47\mb$\cite{Arajs_1960,Gao_2006}.
On the other hand, despite the Curie-Weiss law is not satisfied, roughly estimating the Curie constant
from high-temperature region (2500--4000 K) we find smaller value $\meff\approx4 \mb$,
which approximately equal to the local moment size $\mu_{\rm loc}\approx3.8 \mb$, extracted from the slope of the local susceptibilty
in Sect. II.

\begin{figure}[tbp]
	\includegraphics[angle=-90,width=0.47\textwidth]{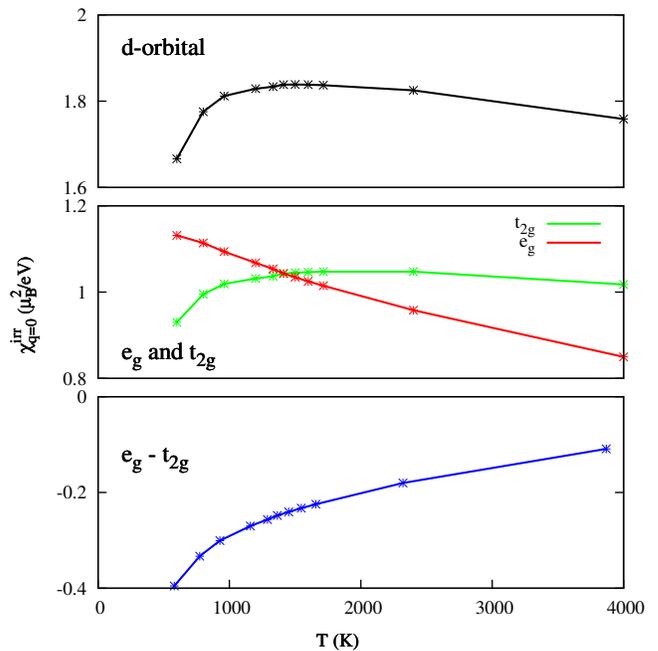}
	\caption{(Color online) Temperature dependence of $\chi^{\rm irr}_{\mathbf{q}=0}$
	         calculated within LDA+DMFT. Top panel --  $\chi^{\rm irr, d}_{\mathbf{q}=0}$,
	         middle panel -- $\chi^{\rm irr, t_{2g}}_{\mathbf{q}=0}$ and
	         $\chi^{\rm irr, e_g}_{\mathbf{q}=0}$, bottom panel -- $\chi^{\rm irr,e_g-t_{2g}}_{\mathbf{q}=0}$. $T=$1290~K, $a=6.91$~a.u.}
	\label{fig:irr_vs_T}
\end{figure}

In order to analyze the role of peculiarities of band structure on non-monotonous temperature behavior of $\chi(T)$,
we calculate $\chi^{\rm irr}_{\mathbf{q}=0}(T)$ projected onto pair sets of orbitals as in Eq.~\ref{eq:decomp}.
The results are shown in Fig.~\ref{fig:irr_vs_T} and inset of the Fig.~\ref{fig:chi_uniform}.
The overall temperature dependence of $\chi^{\rm irr}_{\q=0}(T)$ repeats that of $\chi(T)$,
being however substantially weaker.
The \t2g contribution to $\chi^{\rm irr}_{\q=0}$ has a maximum at the temperature $T\sim 2000$ K,
at which the energy of the thermal fluctuations becomes comparable to the distance of the peak
of the \t2g--projected DOS to the Fermi level, which is about 0.3~eV. The origin of the maximum of $\chi^{{\rm irr}, t_{2g}}_{\q=0}$
is also similar to that, analyzed recently for pnictides\cite{Pniktides}.
The \eg-\t2g contribution has at $T<1000$ K the temperature dependence similar to that of \t2g contribution
but with a negative sign.
The contribution of $e_g$ orbitals decreases almost linearly with increasing temperature.
This is connected with strong (in comparison with $t_{2g}$ orbitals) correlated character of $e_g$ orbitals.
Such a distinct behavior of different orbitals contributions results in the shift of maximum of
total $d$-orbital irreducible susceptibility
to approximately the temperature $T^*$, making it close to the position of uniform susceptibility maximum.
%
The temperature $T^*$ is approximately equal to the characteristic temperature, discussed in Sect. II, above which the
formaltion of local magnetic moments in $\gamma$-iron is expected, explaining naturally a
crossover from Pauli-like to Curie-Weiss-like temperature dependence of the magnetic susceptibility.
The ratio of total uniform susceptibility and irreducible one (Stoner enhancement factor)
at $T\sim 1290$ K is about 10.
It means that ferromagnetic fluctuations, which occur due to proximity of the Fermi level to the peak of
the density of states, are strong in the temperature interval in the vicinity of $T^*$.
Such a large ratio also explains strong temperature dependence of $\chi(T)$ in comparison with
$\chi^{\rm irr}_{\q=0}(T)$.

To estimate exchange interactions we perform the mapping of the considered
electronic system to the effective Heisenberg model. Due to presence of different competing magnetic orders
we consider a rough way to extract the exchange integrals using the electronic properties
in the paramagnetic phase at finite
temperature. To this end we compare a momentum dependence of the
static magnetic susceptibility, $\chi_{\q}$, obtained for the effective Heisenberg model with exchange parameters $J_{\q}$
within the $1/z$-expansion ($z$ is the coordination number) \cite{Izyumov},
\begin{equation}
	\chi_{\q}=\frac{1}{\chi_{\rm loc}^{-1}-J_{\q}/(4\mb^2)},
	\label{eq:CW-Heis}
\end{equation}
with the Eq.~(\ref{eq:chi_DMFT}), which yields
\begin{equation}
	J_{\q}=-4\mb^2\left(\chi^{\rm irr}_{\q}\right)^{-1}+{\rm const}.
	\label{eq:J_ansatz}
\end{equation}
Using the results for $\chi_{\q}^{\rm irr}$ within the LDA+DMFT method
one can obtain the constant in the Eq.~(\ref{eq:J_ansatz})
if one fixes $J_{\q}$ by the condition $\sum_{\q}J_{\q}=0$.
At $T=1290$ K we obtain $J_{\q=0}=\min_{\q}J_{\q}=-2380\, \text{K}$
and $J_{\q=\q_{\rm X}}=\max_{\q}J_{\q}=1172\, \text{K}$.

\section{Conclusions}

We have considered the electronic and magnetic properties of paramagnetic $\gamma$--iron.
The shift of the DOS peak below the Fermi level in $\gamma$-iron causes the
dramatic difference in the electronic and magnetic properties between $\alpha$- and $\gamma$-iron.
The position of this peak is therefore crucial for understanding the magnetic properties
which is similar to recent study of pnictides \cite{Pniktides}.

The account of correlation effects in $\gamma$--iron allows one to conclude that
the effective local moments are formed in this material
at sufficiently large temperature $T>\mbox{1000 K}$ with $\mu_{\rm loc}\approx 3.8 \mb$.
The corresponding inverse local susceptibility $\chi_{\rm loc}^{-1}$ has however apart from the $T$-linear term also
constant contribution, providing strong temperature dependence of the effective local moment
$\mu_{\rm eff}=\sqrt{3 T \chi_{\rm loc}}$,
which  in the temperature range 1200--1400~K is approximately $3 \mb$.
At lower temperatures $\gamma$-iron is found to be
better described in terms of itinerant picture.

The antiferromagnetism of $\gamma$--iron can be understood as occuring due
to band structure features (nesting of some sheets of the Fermi surface, connecting \eg--\eg and \eg--\t2g states).
The obtained antiferromagnetic state with the wavevector close to $(2\pi/a)(1,0,0)$ is found to compete strongly
with the other incommensurate spin-density wave instabilities.
Observed tendency to the magnetic frustration can explain the small N\'eel temperature of
$\gamma$--iron.

The application of obtained results for explaining $\alpha$-$\gamma$ structural transition
in iron and the properties of some iron alloys with fcc structure is of further importance.

The authors are grateful to Yu.~N.~Gornostyrev, A.~V.~Korolev, A.~N.~Ignatenko and
I.~V.~Leonov for useful discussions. This work
was supported by the Russian Foundation for Basic Research (Projects Nos. 13-02-00050, 13-03-00641, 12-02-91371-CT a,
12-02-31207, 11-02-00931-a, 11-02-00937-a, 12-02-31510-mol-a, 10-02-91003-ANF a, the fund of the President of the Russian Federation for the support
of scientific schools NSH-6172.2012.2, the Programs of the Russian Academy of Science: "Quantum microphysics of
condensed matter" (No. 12-P-2-1017, 12-CD-2), "Strongly correlated electrons in solids and structures"  (No. 12-T-2-2001);
the grants of the Ministry of education and science of Russia No. 12.740.11.0026 and 14.A18.21.0076,
the Program of "Dynasty" foundation. Calculation were performed using ``Uran'' supercomputer of IMM UB RAS.

\vspace{1cm}

\section*{Appendix. Relation between $\langle S^2 \rangle$ and the damping $\delta$ of local moments}

In this Appendix we consider the
contribution of the low-frequency part of the local susceptibility
(which is presumably
responsible for contribution of localized degrees of freedom), described by the
 Eq. (\ref{hi_loc}), to the
instaneous local moment. Performing analytical continuation of the Eq. (\ref{hi_loc})
to the imaginary frequency axis with the subsequent summation over Matsubara frequencies, we obtain:%
\begin{align}
\langle(S^{z})^{2}\rangle&  =T\sum_{i\omega_{n}}\chi_{\text{loc}}(i\omega
_{n})=\frac{\mu_{\rm eff}^{2}}{3}\sum_{\omega_{n}}\frac{\delta}{|\omega_{n}|+\delta
}\nonumber\\
&=\frac{\mu_{\rm eff}^{2}}{3}\left\{
1+\frac{\delta }{\pi T}
 \left[\psi \left(n_{\rm m}\right)-\psi \left(1+\frac{\delta }{2 \pi
   T}\right)\right]
\right\}\nonumber\\
&\simeq\frac{\mu_{\rm eff}^{2}}{3}\left[
1+\frac{\delta }{\pi T} \log (n_m) \right]
\label{mom}%
\end{align}
where $n_{\rm m}\sim I/(2\pi T)$ is the largest frequency number, to which the behavior of Eq. (\ref{hi_loc}) extends,
and $\psi$ is the digamma function. It can be also estimated, that the high-energy part of the susceptibility
yields only subleading contribution $O(\delta/(\pi T))$ to the Eq. (\ref{mom}).
In Eq. (\ref{mom}) we can distinguish two regimes.
First, if $\delta\ll \pi T,$ we find $\langle(S^{z})^{2} \rangle \simeq\mu_{\rm eff}%
^{2}/3,$ i.e. the instaneous local moment and the effective moment, extracted from the
Curie law for local susceptibility are close to each other. This is identified with the (sufficiently long-living) local moment regime
in main text. On the other hand,
for $\delta\gtrsim \pi T$ we find $\langle(S^{z})^{2}\rangle \gg \mu_{\rm eff}^2/3,$
which corresponds to the itinerant regime.

\end{document}